\documentclass[runningheads]{svmult}

\usepackage{makeidx}   
\usepackage{graphicx}  
\usepackage{subeqnar}  
\usepackage{multicol}  
\usepackage{physprbb}  
\makeindex             

\begin{document}
\title*{Rapid Growth of Massive Galaxies: A Paradox for Hierarchical
Formation Models}
\toctitle{Rapid Growth of Massive Galaxies: A Paradox for Hierarchical
Formation Models}
%
%
\titlerunning{Rapid Growth of Massive Galaxies}
%
\author{Hsiao-Wen Chen\inst{1}
\and the LCIRS\inst{2} \& GDDS\inst{3} teams}
\authorrunning{Chen et al.}
%
%
\institute{MIT Center for Space Research, Cambridge, MA 02139, USA
\and http://www.ociw.edu/lcirs/lcir.html
\and http://www.ociw.edu/lcirs/gdds.html}

\maketitle              

\begin{abstract}

On behalf of the survey teams I summarize the designs and results of the Las 
Campanas Infrared Survey and Gemini Deep Deep Survey, both of which were 
initiated to understand the nature of red galaxies and to study the history of
stellar mass assembly.  Our results from luminosity function analysis, ISM 
absorption line measurements, and spectral synthesis modeling show that 
near-infrared selected galaxies at $1<z<2$ are not only massive and abundant
but also old and metal enriched, indicating rapid formation of massive systems
at higher redshifts.

\end{abstract}

\section{Background}
  Various galaxy surveys in the optical, near-infrared, and sub-mm have 
uncovered different galaxy populations at redshift $z>2$ (see Steidel, Cimatti,
and Chapman in these proceedings for discussions), but whether these galaxies 
are representative of the galaxy population at high redshifts or how they are 
related to the present-day population is not clear.  For example, rest-frame 
UV-selected samples are presumably sensitive to active star-forming galaxies 
and biased against evolved, quiescent systems that are faint at ultraviolet 
wavelengths, while sub-mm samples select mostly dusty star-forming galaxies.  
In contrast, near-infrared based surveys are sensitive to early-type galaxies 
at $z=0-3$ that have red optical and near-infrared colors and are known to
dominate the total stellar mass at $z=0$.  A complete sample of massive systems
at different epoch allows us to study the history of stellar mass assembly and
offers important clues for discriminating between different galaxy formation 
scenarios (e.g.\ \cite{kc98}).

  But past studies based on near-infrared surveys have yielded inconsistent 
space density measurements of red galaxies at $z\ge 1$ (see \cite{c02} for a 
list of references).  Early-type galaxies are strongly clustered and have no 
prominent narrow-band spectral features at UV wavelengths.  If these red 
galaxies are progenitors of early-type galaxies we see in the local universe 
(rather than dusty star-forming galaxies), then a wide-field infrared survey is
required to minimize the effect of surface density variation between fields due
to strong clustering (e.g.\cite{d00,m01}).  Furthermore, deep galaxy 
spectroscopy is required to obtain a complete sample of these massive systems 
for spectral diagnostics of their ISM and stellar content.

  The Las Campanas Infrared Survey (LCIRS) and Gemini Deep Deep Survey (GDDS) 
are complementary studies designed to probe the nature of red galaxies and to
determine mass assembly history using near-infrared selected galaxies.  In
the following section I summarize the goals and results of the surveys.  

\begin{figure}[ht]
\begin{center}
\includegraphics[width=0.95\textwidth]{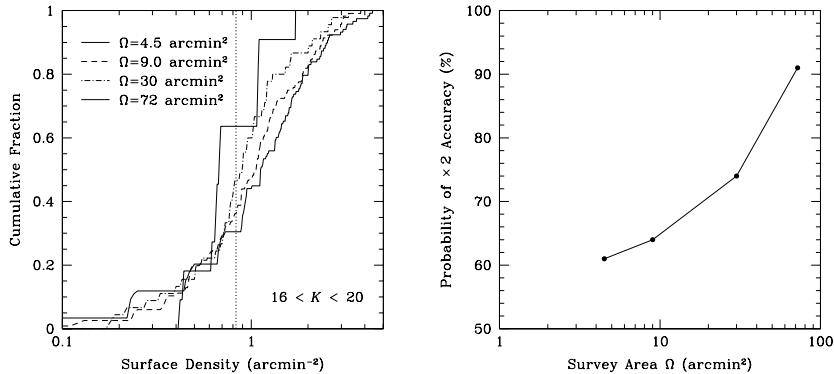}
\end{center}
\caption[]{Uncertainty of surface density measurements of red $I-K\ge 4$ 
galaxies with $16 < K < 20$ in different survey areas $\Omega$.  The surface 
densities were determined empirically within separate sub-areas randomly 
selected from the LCIRS fields.  The left panel shows the cumulative 
probability of different surface density measurements for $\Omega=4.5-72\,{\rm 
arcmin}^2$.  The vertical dotted line indicates the mean surface 
density of 0.83 arcmin$^{-2}$ determined over the entire LCIRS $K$-band 
selected red galaxies.  Based on these curves, we show in the right panel the 
probability of obtaining a measurement within a factor of two accuracy of the 
nominal value versus $\Omega$.}
\label{variance}
\end{figure}

\section{The Las Campanas Infrared Survey}
  The LCIRS is a deep, wide-field near-infrared and optical imaging survey, 
designed to identify a large number of red galaxies at $1<z<2$, while securing 
a uniform sample of galaxies of all types to $z\sim 2$ using photometric 
redshift techniques (\cite{c02,m01,f02}).  The primary objectives are: (1) to 
examine the nature of the red galaxy population and identify evolved galaxies 
at redshifts $z>1$; (2) to study the space density and luminosity evolution of 
early-type galaxies at redshifts $z\le 2$; and (3) to measure spatial 
clustering of massive galaxies, thereby inferring merger rates of these 
galaxies for constraining theoretical models.

\begin{figure}[thb]
\begin{center}
\includegraphics[width=1.0\textwidth]{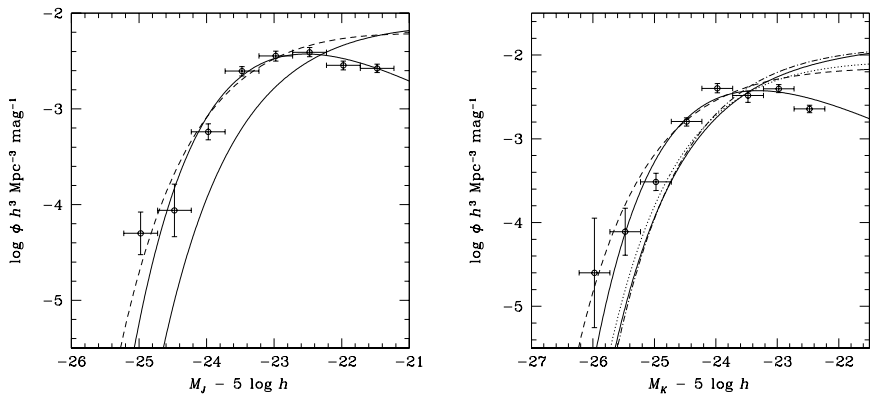} 
\end{center}
\caption[]{Rest-frame $J$ (left) and $K$ (right) galaxy luminosity function of 
$> 4000$ galaxies identified at $1\le z \le 1.5$ in the LCIRS using photometric
redshift techniques.  The points were evaluated using a modified stepwise 
maximum likelihood analysis, explicitly incorporating redshift uncertainties 
from photometric redshift analysis \cite{c03}.  Error bars were estimated using
a bootstrap technique that takes into account photometric uncertainties, 
redshift errors, and sampling biases.  The thick solid curve in each panel 
shows our best-fit Schechter luminosity function.  For comparison, we present 
measurements for $K$-selected galaxies at $0.75 < z < 1.3$ from the K20 survey 
team \cite{p03} (dashed curves).  We also plot local measurements from the 
2MASS survey \cite{c01,k01} (thin curves).}
\label{lf}
\end{figure}

  We have completed the phase-I $H$-band survey to $H=20.5$ over 1.1 ${\rm 
deg}^2$ and are completing the phase-II $K$-band survey to $K=20.6$ over 0.75 
${\rm deg}^2$.  The size of our survey area was determined in order to obtain a
representative measure of the surface density of red galaxies and a significant
signal in clustering analysis \cite{m01}.  Five random fields have been chosen 
and each field covers $\approx 26\times 26\,{\rm arcmin}^2$, corresponding to a
projected co-moving distance 20 $h^{-1}$ Mpc.  This is twice the correlation 
length observed for early-type galaxies at $z=0$ (see \cite{m01} for 
references).  The large survey area of the LCIRS allows us to estimate 
empirically the uncertainties in the surface density measurements of red 
galaxies identified in smaller-area surveys, assuming that the LCIRS galaxy 
sample are representative of these systems at $z\ge 1$.  Figure 1 shows the 
results based on red galaxies with $I-K\ge 4$ found in non-overlapping, random 
sub-areas within the LCIRS fields, in comparison to 0.83 arcmin$^{-2}$ as the 
nominal surface density of $I-K\ge 4$ galaxies determined based on the entire 
LCIRS red sample.  We find based on the curves that a survey area of at least 
70 arcmin$^2$ is necessary to identify a statistically representative sample of
red galaxies.  Our initial clustering analysis based on an $H$-band selected 
sample over 0.62 deg$^2$ of sky shows that red galaxies with $I-H \ge 3$ have a
co-moving correlation length of $9-10\ h^{-1}$ Mpc at $z\simeq 1$, comparable 
to that of present-day early-type galaxies \cite{m01}.
  
  Photometric redshifts have been determined for the entire LCIRS galaxy sample
based on a combination of available optical $BVRIz'$ and near-infrared $JHK$ 
photometry \cite{c03}, making the LCIRS a {\em complete} redshift survey of 
faint galaxies at $z\ge 1$.  But because of relatively large redshift
uncertainties, accounting for these errors in all analyses that make use of 
photometric redshift measurements is necessary to minimize systematic errors.
Adopting an $H$-band selected sample and using a technique that explicitly 
accounts for the {\em non-gaussian characteristics} of photometric redshift 
uncertainties, we have shown that the evolution of the rest-frame $R$-band
co-moving luminosity density $\ell_R$ is characterized by $\Delta\log \ell\,/
\Delta\log (1+z)=0.6 \pm 0.1$ and that luminous early-type galaxies exhibit
only moderate evolution over $z=0.3-1.5$ \cite{c03}.

  Rest-frame $J$- and $K$-band luminosity functions have also been calculated
using $> 4000$ $K$-band selected galaxies at $1\le z\le 1.5$.  Figure 2 shows
the best-fit luminosity functions in $J$ (left panel) and $K$ (right panel)
using modified stepwise maximum likelihood method and STY approach as described
in Chen et al.\ \cite{c03}.  The best-fit Schechter parameters are $M_{J_*} -
5\log\,h = -22.6, \alpha_J = -0.3, \phi_{J_*}=0.0144\,h^3\,{\rm Mpc}^{-3}$, and
$M_{K_*} - 5\log\,h = -23.4, \alpha_K = -0.2, \phi_{K_*}=0.0142\,h^3\,{\rm 
Mpc}^{-3}$.  Figure 2 also shows that our results are consistent at the bright 
end with the estimates presented by the K20 team for 170 galaxies found at 
$0.75 < z < 1.3$ over a total sky area of 52 arcmin$^2$ \cite {p03} (dashed 
curves), but exhibit a hint of a turn-over at the faint-end.  Comparisons with 
existing measurements for local galaxies using 2MASS data \cite{c02,k01} 
indicate that there is approximately $0.4-0.6$ mag fading in luminous ($> L_*$)
near-infrared selected galaxies from $z=1.2$ to $z=0$, the amount of which may
be explained by stellar evolution over the same period of time.  These results
together suggest little/no evolution in the space density of massive early-type
galaxies since $z=1.5$ and support the hypothesis that these systems were 
formed at redshift much beyond $z=1.5$.

\section{The Gemini Deep Deep Survey}

  The GDDS is an ultra-deep spectroscopic survey of near-infrared selected
galaxies at $0.7<z<1.8$ \cite{a03,s04}.  The primary objectives are: (1) to 
obtain spectral diagnostics of the ISM and stellar content of massive galaxies 
at high redshift; (2) to construct the stellar mass function over the target 
redshift range; and (3) to determine ages and star-formation histories of 
evolved systems at $z>1$.

  The survey makes use of nod \& shuffle observing mode \cite{gb01} for 
accurate sky subtraction and fringe removal at wavelength beyond 7500 \AA.  It
is aimed at identifying near-infrared selected faint galaxies at $z>0.8$ to a 
high completeness level.  We have observed 301 spectra in four fields using 
GMOS on Gemini North.  Each of the four fields covers $30\,{\rm arcmin}^2$ 
field of view and is located in a separate LCIRS field.  The pointing and 
target selection of the GDDS sample is guided by known information in the 
wider-field LCIRS sample to maximize the efficiency of identifying a 
representative sample of faint galaxies at $z>0.8$.  For example, each GDDS 
pointing is selected from within an LCIRS field to have the number of red 
galaxies comparable to the global mean, which is important for reliable volume 
density analyses of early-type galaxies.  In addition, we exclude foreground 
contaiminating sources at $z<0.8$ using photometric redshifts available for all
LCIRS sources.  Figure 3(a) shows a comparison of the observed $V-I$ color 
versus $z$ for CFRS \cite{l95}, GDDS \cite{a03}, and UV-selected galaxies 
\cite{m03}.  It demonstrates that the GDDS is probing a representative galaxy 
population concentrated at $0.8 \le z <2$ with a broad-range of star formation 
history, from quiescent through active star-forming systems.  Conversely, the 
GDDS results offer necessary tests and verification of the accuracy of LCIRS 
photometric redshifts.  Figure 3(b) shows that the rms scatter between $z_{\rm 
phot}$ and $z_{\rm spec}$ is $\sigma_{\Delta z/(1+z)}\approx 0.1$, consistent 
with previous results based on a smaller number of galaxies at $0\le z <1$ 
\cite{c03}.

\begin{figure}[thb]
\begin{center}
\includegraphics[width=1\textwidth]{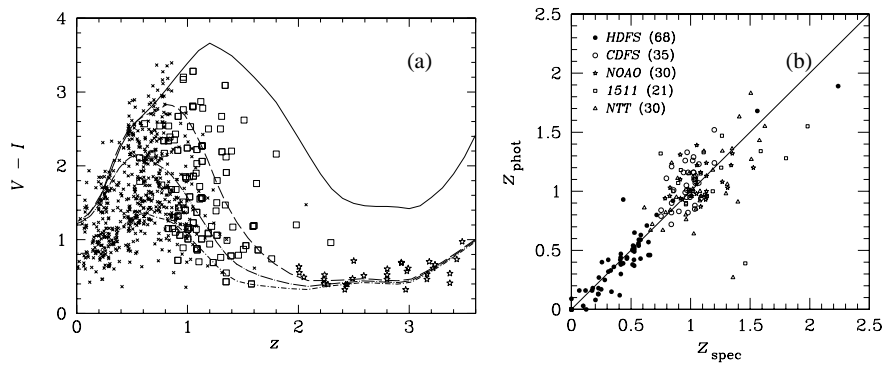}
\end{center}
\caption[]{(a) The observed $V-I$ color versus redshift for the GDDS sample 
(squares), in comparison to the CFRS sample \cite{l95} (crosses), UV-selected 
galaxies \cite{m03} (stars), and model predictions for a non-evolving 
elliptical galaxy (solid curve), to an exponentially declining star formation 
rate (SFR) with e-folding time of $\tau=1$ (dashed curve) and 2 Gyr (long 
dash-dotted curve), and a constant SFR (short dash-dotted curve).  (b) 
Comparison of photometric and spectroscopic redshifts for $\sim 200$ galaxies 
identified in the LCIRS.  Spectroscopic redshifts for galaxies in the HDFS and 
CDFS were collected in part from the literature and in part from our own 
observations.  Spectroscopic redshifts for galaxies in the NOAO, 1511, and NTT 
fields were obtained from the GDDS.}
\label{cz}
\end{figure}

  The GDDS data have revealed a wealth of information regarding the ISM and
stellar content of massive galaxies at $z > 1$.  Sample spectra are presented
in Figure 4, where we show various GDDS composite spectra as well as individual
sample spectra to demonstrate the data quality.  The full sample of GDDS 
spectra is presented in Abraham et al.\ \cite{a03}.  Specifically, we show in
panel (a) the composite spectrum of early-type (absorption-line dominated) 
galaxies, panel (b) intermediate-type galaxies, and in panel (c) late-type 
(emission-line dominated) galaxies.  For comparison, we include in panel (a) 
the SDSS composite for luminous red galaxies (LRG) at $\langle z\rangle=0.3$ 
\cite{e03} (red curve).  Our composite spectrum for early-type galaxies at 
$z=1-2$ is strikingly similar to the LRG composite spectrum at $\langle z
\rangle=0.3$, showing little evolution in the stellar content of massive and 
quiescent systems over the redshift interval spanned by the two samples.  We 
also show the spectrum of a galaxy at $z=1.34$ in panel (d) together with the
spectrum of the host galaxy of 53W091 at $z=1.55$ \cite{d96} (red curve) and a 
starburst galaxy at $z=1.67$ in panel (e).  These spectra demonstrate that we 
have good signal-to-noise ratios to obtain accurate and precise redshifts of 
individual galaxies using both rest-frame ultraviolet metal absorption features
and weak continuum bumps originating in evolved stellar populations.

\begin{figure}[!]
\begin{center}
\includegraphics[width=1.0\textwidth]{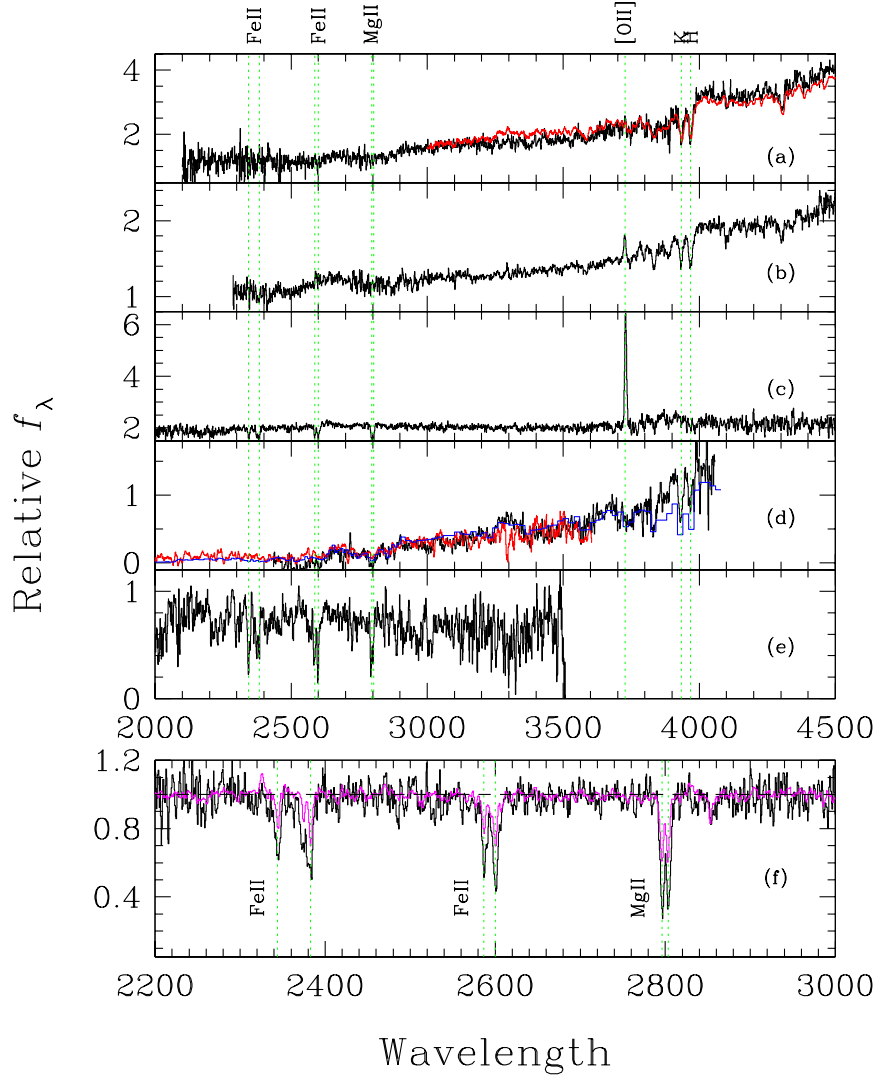}
\end{center}
\caption[]{(a)-(c) GDDS composite spectra for respectively early-type, 
intermediate, and late-type galaxies at $z=1-2$.  (d) The spectrum of a galaxy 
at $z=1.34$ from the GDDS, together with the spectrum of the host galaxy of 
53W091 at $z = 1.55$ \cite{d96} in red, and the model spectrum of a single 
burst system formed at $z_f=6$ in blue.  (e) The spectrum of a starburst galaxy
at $z = 1.67$, showing strong metal absorption features.  (f) A normalized 
composite spectrum of 13 massive star-forming galaxies identified at $\langle z
\rangle = 1.6$ from the GDDS, together with a normalized composite spectrum of 
local starburst galaxies \cite{t03}.}
\label{fe2}
\end{figure}

\begin{figure}[tbh]
\begin{center}
\includegraphics[width=1.0\textwidth]{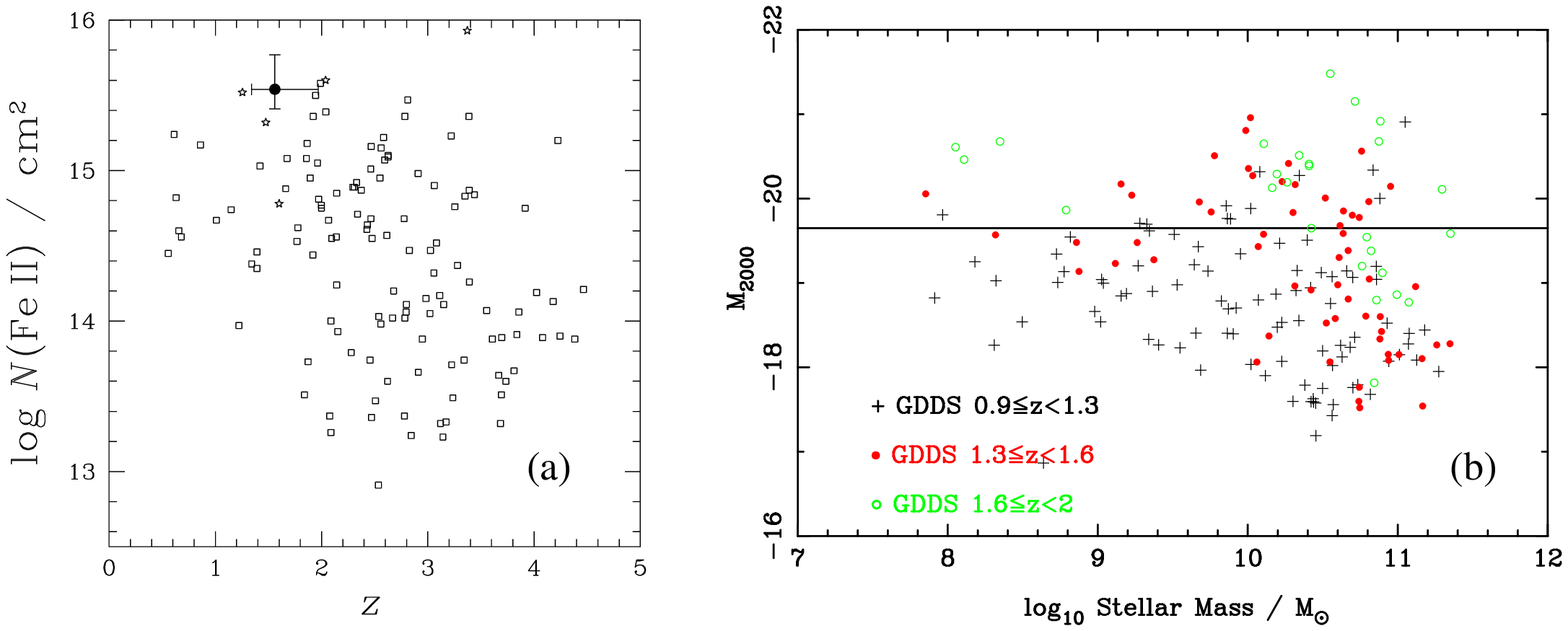}
\end{center}
\caption[]{(a) Fe\,II column density vs.\ redshift in different systems: solid 
point with error bars represents the GDDS measurement; open squares are for 
damped Ly$\alpha$ absorption systems compiled from the literature; and stars 
are for the ISM of gamma-ray burst host galaxies.  (b) Rest-frame UV magnitude 
versus inferred stellar mass for galaxies in three redshift intervals.  The 
solid line indicates the minimum UV flux of the galaxies included for the
composite spectrum presented in panel (f) of Figure 4.}
\label{fe2}
\end{figure}

 In addition, we have constructed a composite spectrum for 13 $K$-band 
selected galaxies at $1.3 < z < 2$ from the GDDS sample that are also luminous 
in the UV with $M_{2000}<-19.5$ \cite{s04}.  A normalized version is shown in
panel (f) of Figure 4, together with a normalized composite spectrum of local 
starburst galaxies from Tremonti et al.\ \cite{t03}.  The composite spectrum 
shows prominent absorption features from Mg\,II, Fe\,II together with Mg\,I and
Mn\,II, all of which are stronger than those observed in local starburst 
systems.  It indicates that the ISM of these massive star-forming galaxies has
been heavily enriched with metals by $\langle z\rangle=1.6$.  Based on a 
curve-of-growth analysis of all available transitions, Savaglio et al.\ 
(\cite{s04}) find a mean Fe\,II column density of $\log\,N_{\rm Fe\,II}/({\rm 
cm}^2)=15.54_{-0.13}^{+0.23}$ that is among the highest observed in either 
known damped Ly$\alpha$ absorption systems or in the ISM of host galaxies of 
gamma-ray bursts (Figure 5a).  While the metallicity is uncertain due to the
unknown neutral hydrogen content, it is clear that the ISM of these galaxies 
contains abundant heavy elements.  Because of the unknown nature of the 
galaxies that give rise to damped Ly$\alpha$ systems, direct column density 
measurements of the ISM of luminous galaxies at high redshift offer a promising
alternative for understanding cosmic chemical enrichment history.  

  We also determine the formation epoch of the stars seen in quiescent systems 
that show no signs of recent star formation (lack of emission features in the
rest-frame UV).  Using a likelihood analysis, we derive a minimum age for each 
system by comparing predicted spectral energy distributions (SEDs) from 
different stellar synthesis models with the observed ones established from the 
GDDS spectra and available LCIRS optical and near-infrared photometry.  An 
example is shown in panel (d) of Figure 4 for a red galaxy at $z = 1.34$.  The
model (blue curve) was obtained for a single burst occurring at $z_f=6$ and
passively evolving to lower redshift.  The agreement between the observed
spectrum and models suggests that this galaxy is at least 3.5 Gyr old, pushing
the formation epoch of this galaxy to beyond $z_f=6$ (McCarthy et al.\ 2004).  

  Finally, we estimate the total stellar mass of every galaxy in the GDDS 
sample using a $K$-band mass-to-light ratio determined from comparing the 
observed multi-color data with a grid of model SEDs with varying dust 
extinction, metallicity, and star formation history.  Figure 5(b) shows 
rest-frame UV flux (expressed in $M_{2000}$) versus the inferred stellar masses
of individual galaxies in three redshift bins spanned by the GDDS sample.  We 
see that a large fraction of the GDDS galaxies have masses $> 
10^{10}\,M_\odot$, and these systems exhibit on average increasing rest-frame 
UV fluxes (and therefore increasing star formation rate) with increasing 
redshift.  In addition, while few systems with stellar masses 
$< 10^{10}\,M_\odot$ are seen at $z>1.3$ in the GDDS sample due to the $K$ 
limit of the survey, a tendency for the most massive galaxies to be UV 
quiescent is found in the low-redshift sample (crosses)--a trend which is
similar to what is seen at $z=0$.  This result supports the view that most
star formation takes place in lower mass systems even at $z>1$ (Glazebrook et 
al.\ 2004).

  In summary, the wide-field multi-color imaging data from the LCIRS and the
high-quality nearly complete spectral sample from the GDDS together allow us to
conduct a comprehensive study of massive early-type galaxies at $z=1-2$.  Our
luminosity function analysis indicates that most and perhaps all of the 
present-day infrared luminous galaxies were in place by $z\approx 1.2$.  The
rest-frame UV spectra of massive star-forming systems show abundant metals that
arise in the ISM of these galaxies, indicating an early heavy element 
enrichment by $z\approx 1.6$.  Finally, spectral synthesis modelling showed 
that near-infrared selected galaxies at $z>1$ are massive and that the 
quiescent ones with no evidence of recent star formation are genuinely old and 
likely to form at $z>4$.  Together these results suggest a rapid and early 
formation process of massive galaxies, presenting difficulties for 
conventiional hierarchical formation scenarios.

%

\end{document}